\documentclass[a4paper,nocompress]{spie}  %
\pdfoutput=1
\usepackage{amsmath,amsfonts,amssymb}
\usepackage{graphicx}
\usepackage[colorlinks=true, allcolors=blue]{hyperref}

\usepackage[T1]{fontenc}      %

\graphicspath{{./figures/}{./}}       %
\DeclareGraphicsExtensions{.pdf,.png,.jpg}

\usepackage[squaren,thinspace,thickqspace]{SIunits}
  \addunit{\photons}{photons}
  \addunit{\mas}{mas}
  
  \newcommand{\arcsec}{\arcsecond}

\usepackage{color}

\usepackage{soul}

\usepackage{verbatim}

\newcommand{\V}[1]{{\boldsymbol{#1}}}                 %
\newcommand{\M}[1]{{\mathbf{#1}}}                     %
\newcommand{\T}{^{\mathrm{T}}}                        %
\newcommand{\Inv}{^{-1}}                              %
\newcommand*{\Expect}{\mathbb{E}}
\newcommand{\Norm}[1]{\left\Vert #1\right\Vert}       %
\newcommand{\Var}{\operatorname{Var}}                 %
\newcommand{\estim}[1]{\widehat{#1}}                  %

\newcommand*{\Tag}[1]{\mathsf{#1}}

\usepackage{xspace}
\newcommand{\cf}{\emph{cf.}\xspace}      %
\newcommand{\eg}{\emph{e.g.}\xspace}     %
\newcommand{\etal}{\emph{et al.}\xspace} %
\newcommand{\ie}{\emph{i.e.}\xspace}     %

\newcommand{\refeq}[1]{Eq.~(\ref{#1})}
\newcommand{\reffig}[1]{Fig.~\ref{#1}}
\newcommand{\refsec}[1]{Sec.~\ref{#1}}

\newcommand{\tmpA}{}
\newcommand{\tmpB}{}

\title{The wavefront sensing making-of for THEMIS solar telescope}

\author[a]{Michel Tallon}
\author[a]{\'Eric Thi\'ebaut}
\author[a]{Maud Langlois}
\author[b]{Bernard Gelly}
\author[b]{Richard Douet}
\author[c]{Cl\'ementine B\'echet}
\author[d]{Lo\"{\i}c Denis}
\author[a]{Gil Moretto}

\affil[a]{Univ Lyon, Univ Lyon1, Ens de Lyon, CNRS, Centre de Recherche Astrophysique de
Lyon UMR5574, F-69230, Saint-Genis-Laval, France}

\affil[b]{CNRS UPS3718 - THEMIS, Via Lactea s/n, ES-38205 La Laguna, Canary
Islands, Spain}

\affil[c]{Institute of Mathematical and Computational Engineering, Pontificia
Universidad Cat\'olica de Chile, Santiago, Chile}

\affil[d]{Univ. Lyon, UJM-Saint-\'Etienne, CNRS, Institut d'Optique Graduate
School, Laboratoire Hubert Curien UMR 5516, F-42023, Saint-\'Etienne, France}

\authorinfo{Send correspondence to Michel Tallon, E-mail: 
mtallon@obs.univ-lyon1.fr}

\begin{document} 
\maketitle

\begin{abstract}

An adaptive optics system with a single deformable mirror is being implemented
on the THEMIS 90cm solar telescope. This system is designed to operate in the
visible and is required to be as robust as possible in order to deliver the
best possible correction in any atmospheric conditions, even if wavefronts are
sensed on some low-contrast solar granulation. In extreme conditions, the
images given by the subapertures of the Shack-Hartmann wavefront sensor get
randomly blurred in space, in the set of subapertures, and the distribution of
blurred images is rapidly changing in time, some of them possibly fading away.
The algorithms we have developed for such harsh conditions rely on inverse
problem approach. As an example, with the gradients of the wavefronts, the
wavefront sensor also estimates their errors, including their covariance. This
information allows the control loop to promptly optimize itself to the fast
varying conditions, both in space (wavefront reconstruction) and in time. A
major constraint is to fit the calculations in a low-cost multi-core CPU. An
overview of the algorithms in charge of implementing this strategy is
presented, focusing on wavefront sensing.

\end{abstract}


\section{Introduction}
\label{s:intro}

Adaptive optics (AO) is spreading among solar
telescopes\cite{Rimmele_&_Marino_2011}. In this context, an adaptive optics
system is being installed on THEMIS\footnote{T\'elescope H\'eliographique pour
l'Etude du Magn\'etisme et des Instabilit\'es Solaires, \ie Heliographic
Telescope for the Study of Magnetism and Solar Instabilities} 90 cm solar
telescope at Teide Observatory, Tenerife, Canaria Islands. This new
developpement offers the opportunity to implement new methods of reconstruction
and control, driven by the requirement that the AO runs unsupervised at its
best performance in any atmospheric conditions. The AO system is combined with
a major refurbishment of the telescope optics, from the telescope secondary
down to the spectrograph entrance, in order to keep the unique
spectropolarimetric capabilities of this telescope\cite{Gelly_et_al_2016}.
These changes in the optical path should lead to an AO corrected telescope fit
for high quality polarimetry.

\begin{figure}[t]
  \hfill
  \centering
  \begin{minipage}[b]{0.4\textwidth}
    \centering
    \includegraphics[width=0.6\textwidth]{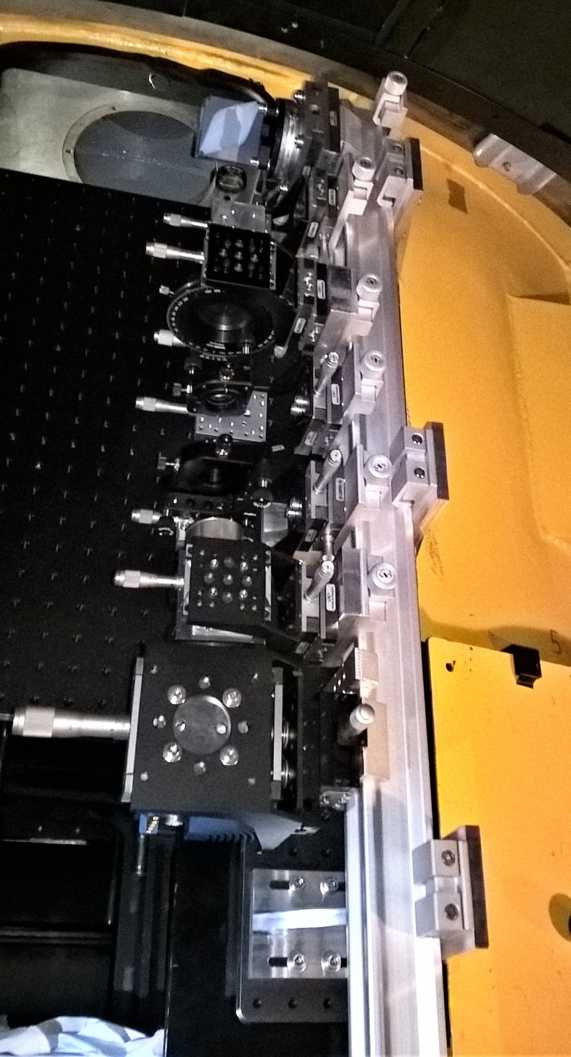}
    \caption{Shack-Hartmann wavefront sensor installed at THEMIS telescope.}
    \protect\label{f:pupil}
  \end{minipage}
  \hfill
  \begin{minipage}[b]{0.4\textwidth}
    \centering
    \includegraphics[width=0.75\textwidth]{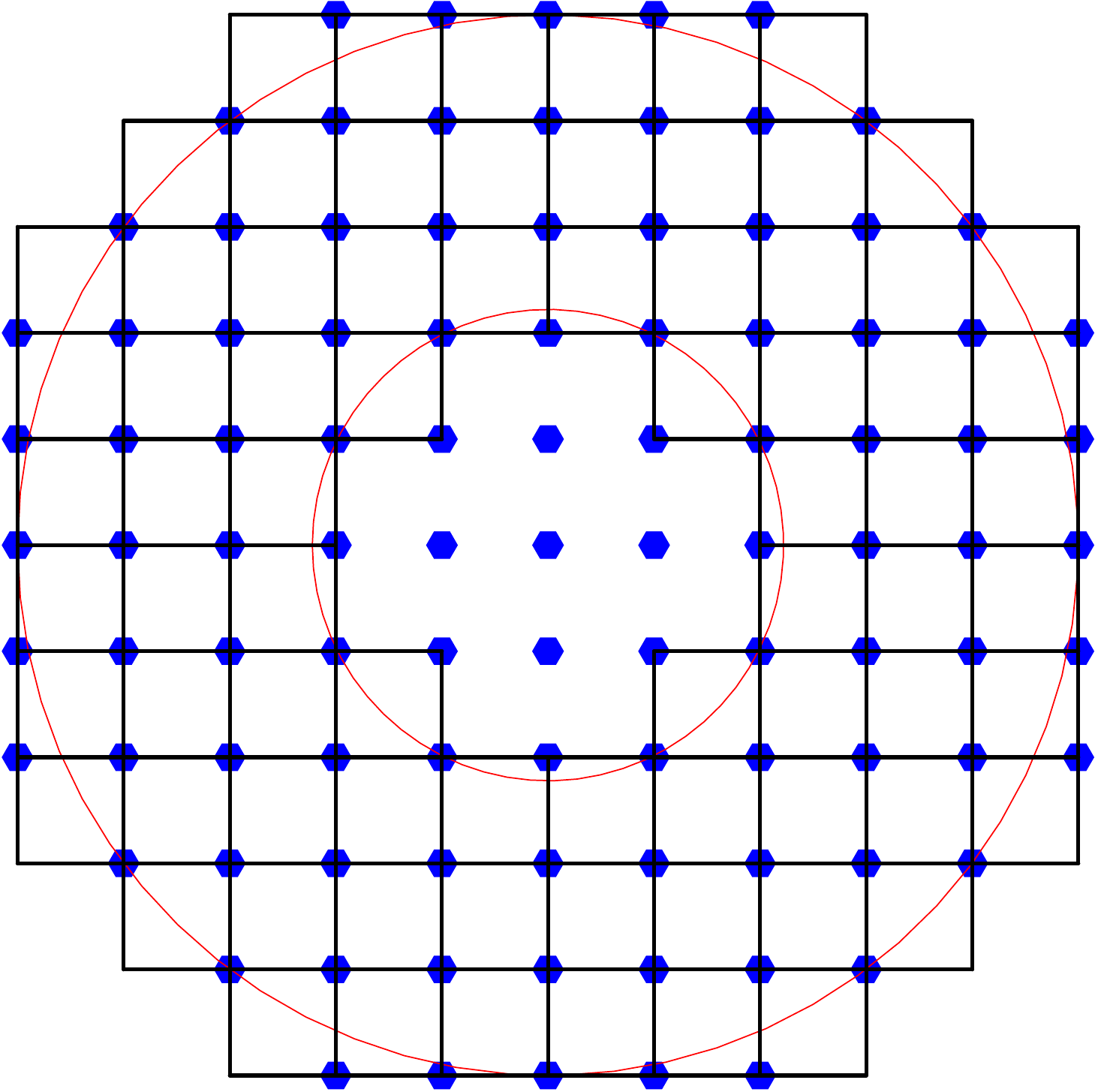}
    \caption{Geometry of the THEMIS adaptive optics system, with a $10\times10$
    Shack-Hartmann wavefront sensor (76 active subapertures) and a ALPAO DM97
    deformable mirror with 97 actuators. The red circles denote the shape of
    the telescope pupil with its central obscuration.}
    \protect\label{f:themis_AO_geometry}
  \end{minipage}
  \hspace*{\fill}\par
\end{figure}

As shown by figure~\ref{f:themis_AO_geometry}, the AO system is based on a
deformable mirror with 97 actuators ($11\times11$), from ALPAO, and a
Shack-Hartmann wavefront sensor with 76 subapertures ($10\times10$) with a
Fried geometry\cite{Fried_1977} (\reffig{f:wfs}). In order to keep the cost as
low as possible, the Real Time Computer is a single PC with 4 cores at 4\,GHz,
and run the control loop at 1~kHz.

This paper will focus on the new wavefront sensing method at the root of the
control loop. The method aims at measuring optimally the displacements of the
sub-images delivered by the wavefront sensor on the solar granulation seen with
~2\% contrast and a digitization on 8 bits only.

Since the day time atmospheric conditions may be quite instable, with
sub-images randomly blurred in space and in time, some sub-images even fading
away, the proposed method also estimate the errors corresponding to the
measurements obtained at each frame, including their covariances. These errors
will allow wavefront reconstruction to optimally take into account the
variations of the errors in the pupil and will allow the loop to optimize
itself to fast varying conditions.

Since the measurements now come with their associated precision matrix (invert
of the covariance matrix) in real time, the computation of the commands for the
adaptive optics correction must also be able to take them into account. The
usual method relies on a command matrix obtained with a pseudo inverse of the
interaction matrix. This inversion is usually computed with truncated SVD and
may include modal filtering\cite{Gendron_&_Lena_1994}. Such an estimator may
include fixed weights, for instance to take laser guide star elongation into
account\cite{Tallon_et_al_2008}. But here, this scheme would implies to compute
a new command matrix at each frame since precision matrix changes at this rate.
The solution is to compute the commands by an iterative method, without
matrices\cite{Thiebaut_&_Tallon_2010}, since it only uses the direct model
(sparse interaction matrix, weights, and priors) that can be changed on the
fly. The algorithm to compute the commands from the wavefront sensor
measurements is out of the scope of this paper where we focus on wavefront
sensing on open-loop real data acquired with THEMIS wavefront sensor.

The wavefront sensing method presented in the following is based on the fit of
a reference image used as a model, on the pixels of the sub-images. The
covariance matrix of the measurements can be obtained if the error on the pixel
values is known. So the method starts with a specific calibration of the
detector in order to estimate the error of each pixel value.

The next sections will overview successively the pre-processing of the pixels,
with a particular step to compute an estimate of the error on each pixel value,
the estimation of the slopes and their covariances from the pre-processed
pixels, and the estimation and the update of the reference image used as a
model for the estimation of the slopes.

\section{Pre-processing of raw pixels}
\label{s:preprocessing}

Pixel values are at the root of the whole processing. In order to estimate the
covariance of the errors on the slope measurements, we need to estimate the
variances of the pixel values. This implies an additional step to the usual dark /
flat correction process, which is presented in this section. For each pixel
$i$, we seek for the incident flux in arbitrary unit and the standard deviation of the errors on the estimated
flux. In order to simplify the derivation of the equations, we assume here that
all the necessary calibration frames (see after) are available with the same
exposure time as the data (wavefront sensor frames). This assumption is quite
practicable here since the flux from the Sun is fairly stable. With this
assumption, a pixel raw value in ADU, $r_i$, can be written as:
\begin{equation}%
  r_i = \frac{t_i\,\phi_i}{g_i} + b_i + n_i,
\end{equation}
where $t_i$ is the transmission from the sky down to the pixel $i$, including
the quatum efficiency of the detector, $\phi_i$ is the flux of the source in an
arbitrary unit proportional to photons, $g_i$ is the gain (e$^-$/ADU), and
$b_i$ and $n_i$ are respectively the bias and the noise, both in ADU.

We use tree types of calibration frames obtained with the same exposure time:
\begin{itemize}

  \item \emph{dark frames}: $\phi_i = 0$, \ie the pixel does not receive any
  flux. A large set of dark frames are recorded to compute the mean and
  variance on each pixel, denoted $\Expect(r_i^\Tag{dark})$ and
  $\Var(r_i^\Tag{dark})$ respectively.

  \item \emph{flat frames}: $\Expect(\phi_i) = \Expect(\phi^\Tag{flat})$, \ie
  we assume that all the pixels receive the same flux on average, even if this
  flux varies in time. This flat frame is obtained on the Sun while the
  telescope is speedily wandering other the solar granulation. We only need the
  mean of a large set of frames, denoted $\Expect(r_i^\Tag{flat})$.

  \item \emph{static frames}: $\Var(t_i\,\phi_i) = \Expect(t_i\,\phi_i)$, \ie
  we can assume Poisson statistics of the counts, which holds even if the
  pixels receive different amounts of light. In this case, we use an internal
  stabilized source that shed some, non-uniform but stable in time, light on
  the detector (a Mikrotron EoSens 4CXP camera). Here again, a large set of
  frames is used to compute the empirical mean and variance on each pixel,
  denoted $\Expect(r_i^\Tag{stat})$ and $\Var(r_i^\Tag{stat})$ respectively.

\end{itemize}

Compared to the usual detector calibration, we thus need the variance of the
\emph{dark frames}, and a set of so-called \emph{static frames}. In principle,
a source could deliver both \emph{flat} and \emph{static} frames, but the
``flattening method'' generally introduces fluctuations of the source, as it is
the case with solar granulation.

The \emph{flat frames} are obtained through the field-stop of the wavefront
sensor; so their mean is also used to determine the regions of interest (ROI)
on the detector for each sub-image. The flat correction also account for the
non-uniform transmission of the optics (\eg vignetting by the microlenses).

The calibrated value of the pixel $i$ is obtained from a linear combination 
of raw value $r_i$, as usual\cite{Lena_et_al_2012}:
\begin{equation}%
  d_i = \alpha_i (r_i - \beta_i).
\end{equation}
By assuming $\Expect(d_i) = \Expect(\phi_i)/\Expect(\phi^\Tag{flat})$, we get:
\begin{align}
  \alpha_i &= \frac{1}{\Expect(r_i^\Tag{flat})-\Expect(r_i^\Tag{dark})}, \\
  \beta_i &= \Expect(r_i^\Tag{dark}).
\end{align}
Thus if the flat frame is obtained on the Sun itself, the calibrated values of
the pixels are around unity.

Assuming that the value $d_i$ is an estimation of the expected value (we have a
single sample for one pixel!), the variance of the pixel value is also obtained
from a linear combination:
\begin{equation}%
  \Var(d_i) \approx \frac{\max(d_i,0) + v_i}{u_i},
\end{equation}
with
\renewcommand*{\tmpA}[1]{\Expect(r_{i}^\Tag{#1})}
\renewcommand*{\tmpB}[1]{\Var(r_{i}^\Tag{#1})}
\begin{align}%
  g_i &= \frac{\tmpA{stat}-\tmpA{dark}}{\tmpB{stat} - \tmpB{dark}}, \\
  u_i &= g_i\left(\tmpA{flat}-\tmpA{dark}\right), \\
  v_i &= g_i\ \frac{\tmpB{dark}}{\tmpA{flat}-\tmpA{dark}}.
\end{align}

The estimation of the errors on the pixels thus needs the additional
computation of $\Var(r_i^\Tag{dark})$, $\Expect(r_i^\Tag{stat})$ and
$\Var(r_i^\Tag{stat})$.

In the following, we gather all the pixels $d_i$ of a sub-image $k$ in a single
vector $\V{d}_k$. In the same way, their corresponding weights $1/\!\Var(d_i)$
are gathered in the precision matrix $\M{W}_k$ of sub-image $k$. Since the
covariance terms are neglected, the matrix $\M{W}_k$ is diagonal.

\section{Measurement with a known reference image}
\label{s:measurement}

\begin{figure}[t!]
  \centering
  \includegraphics[width=0.75\textwidth]{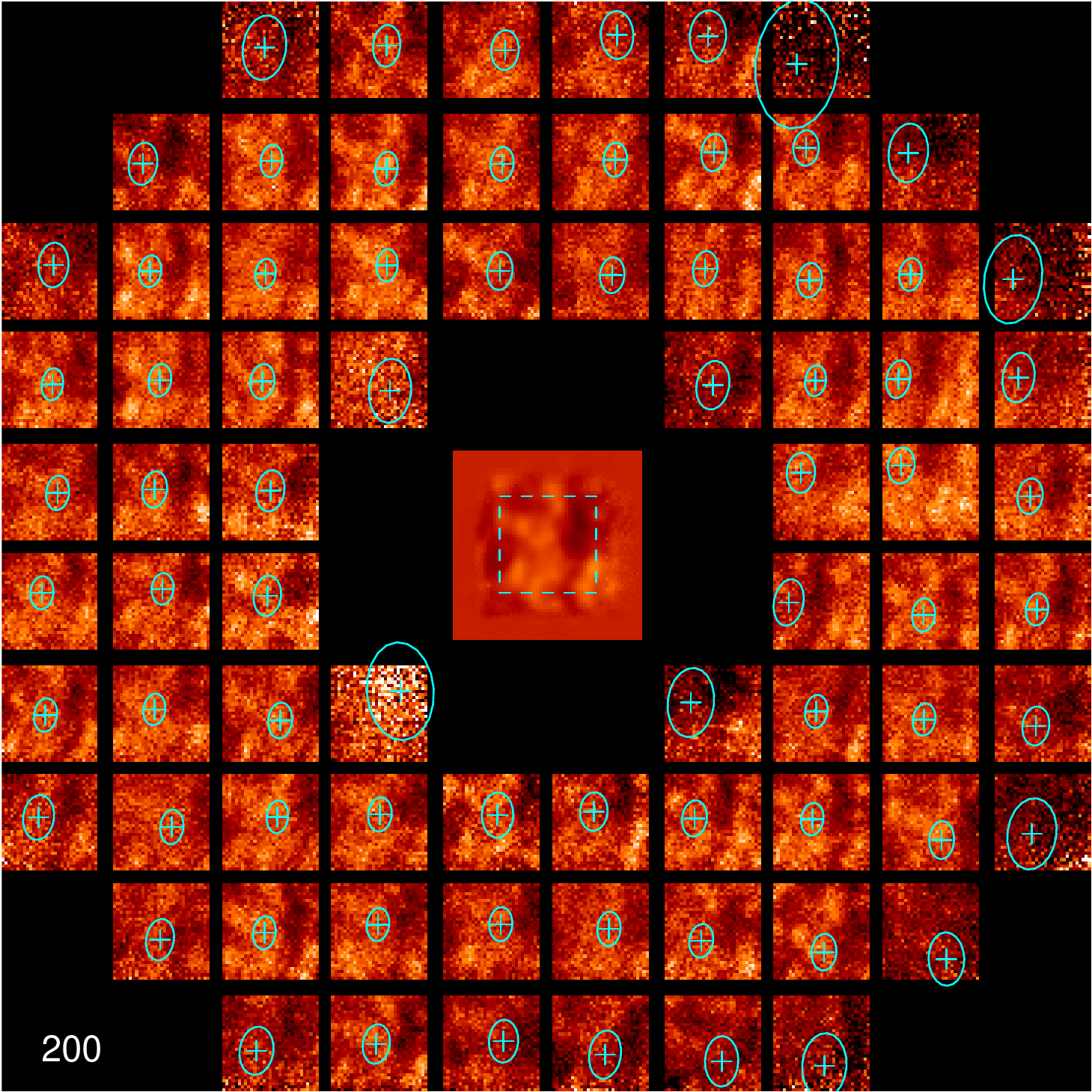}
  \caption{Superimposition of a wavefront frame pre-processed as explained in
  \refsec{s:preprocessing}, and the result of the measurement process as
  explained in \refsec{s:measurement}. The 76 sub-images show a $10\arcsec$
  field-of-view on $30\times30$ pixels, obtained with a $\sim200\micro\second$
  exposure time at $1\kilo\hertz$. The contrast of the granulation seen in each
  subaperture is $\sim2\%$. The crosses show the estimated locations of the
  sub-images for this frame, and the ellipses the estimated standard deviation
  of these locations. This frame shows that the accuracy along the x-axis is
  better than the one along the y-axis, and that subapertures at the edges of
  the pupil are less accurate. The reference image shown at the center is
  estimated from the previous frames on a larger field-of-view (dashed square)
  and with more details than the field-of-view of the subapertures (\cf
  \refsec{s:update}).}
  \protect\label{f:wfs}
\end{figure}

Assuming Gaussian noise, maximum likelihood amounts to minimizing the following
objective function:
\begin{equation}\label{e:psi_cost}
  \psi(\V\theta) = \sum_{k=1}^{n}\left\{
     \eta_k \,\Norm{\V{d}_k - \alpha_k\,\M{R}_k(\V{s}_k).\V{r}}_{\M{W}_k}^2
      - m_k\,\log\eta_k\right\} + \mu \Norm{\M{D}.\V{r}}_2^2
\end{equation}
with $\V{d}_k$ the calibrated pixel values in sub-image $k$, $\M{W}_k$ the
corresponding precision matrix, $m_k$ the number of pixels in the sub-image
$k$, $\M{R}_k(\V{s}_k)$ an operator that shifts the \emph{reference image}
$\V{r}$ by an offset $\V{s}_k$, $\alpha_k$ a rescaling of the reference image
for modeling the scintillation and $\eta_k$ a rescaling of $\M{W}_k$ to take
modeling errors into account. Thus we aim here at fitting each sub-image
$\V{d}_k$ with a known \emph{reference image}, $\V{r}$, that can be shifted by
an offset $\V{s}_k$ and rescaled by a factor $\alpha_k$. Because of
registration errors, we expect that sub-images do not perfectly match the
reference image, for that reason we allow the precision matrix to be rescaled
by a quality factor $\eta_k$ for each sub-image. As the quality factors are
part of the parameters to fit, the associated logarithm term must be kept in
the objective function.

In \refeq{e:psi_cost}, the term $\mu \Norm{\M{D}.\V{r}}_2^2$ is meant to
regularize the estimation of the reference image $\V{r}$ that will be addressed
in the next section. In this section, we assume the reference image $\V{r}$
fixed and known, this term has no effect.

We are only looking for the offsets $\V{s}_k$ and their covariances, but need
to fit in real time all the parameters $\V{\theta} = \{\eta_k, \alpha_k,
\V{s}_k\}_{k=1,...,n} \cup \V{r}$.

The operations applied to the reference image are non linear with respect to
the shifts $\V{s}_k$. A linearization of $\alpha_k\,\M{R}_k(\V{s}_k).\V{r}$ is
necessary to derive an algorithm fast enough to be applied in real time. This
linearization step relies on a first order expansion and a change of variable
around an expected value of $\V{s}_k$, as detailed in Thi\'ebaut
\etal\cite{Thiebaut_et_al_2018}. The likelihood term for one sub-image now
writes:
\begin{equation}\label{e:phi_cost}
  \varphi_k(\V{u}_k) = \Norm{\V{d}_k - \M{H}_k.\V{u}_k}_{\M{W}_k}^2
                     = \gamma_k - 2\,\V{b}_k\T\!\!.\V{u}_k
                       + \V{u}_k\T\!\!.\M{A}_k.\V{u}_k,
\end{equation}
with
\newcommand{\ArrayWithDelimiters}[4]{%
  \left #1\begin{array}{#2}#4\end{array}\right #3}
\newcommand{\Vector}[1]{\ArrayWithDelimiters{\lgroup}{c}{\rgroup}{#1}}
\begin{equation}%
  \V{u}_k = \alpha_k\,\Vector{1 \\ x_k \\ y_k}
\end{equation}
where $(x_k, y_k)$ is the sought relative displacement from the predicted
position, $\M{H}_k$ is a $m_k\times3$ matrix obtained from a first order
expansion\cite{Thiebaut_et_al_2018} of the shifted reference image, and:
\begin{align}%
  \gamma_k &= \Norm{\V{d}_k}_{\M{W}_k}^2, \\
  \V{b_k}  &= \M{H}_k\T\!.\M{W}_k.\V{d}_k, \\
  \M{A}_k  &= \M{H}_k\T\!.\M{W}_k.\M{H}_k.
\end{align}

The solution of \refeq{e:phi_cost} is:
\begin{equation}%
  \estim{\V{u}}_k = \M{A}_k\Inv\!.\V{b}_k,
\end{equation}
and the corresponding value of $\varphi_k(\V{u}_k)$ at the minimum is:
\begin{equation}\label{e:phi}
  \varphi_k(\estim{\V{u}}_k) = \gamma_k - \,\V{b}_k\T\!.\V{\estim{u}}_k
\end{equation}

The next step is to estimate the rescaling of the weights $\M{W}_k$ from
\refeq{e:psi_cost}. After this linearization and solving for $\estim{\V{u}}_k$
for each sub-image $k$, \refeq{e:psi_cost} now writes:
\begin{equation}%
  \psi(\V\theta) = \sum_{k=1}^{n}\left\{
                \eta_k \, \varphi_k(\estim{\V{u}}_k) - m_k\,\log\eta_k\right\},
\end{equation}
the value of the likelihood term, $\varphi_k(\estim{\V{u}}_k)$, being known
from \refeq{e:phi}. Minimizing $\psi(\V\theta)$ for each sub-image, yields:
\begin{equation}%
  \eta_k = \frac{m_k}{\varphi_k(\estim{\V{u}}_k)}.
\end{equation}

For each sub-image $k$, the covariance matrix, $\M{C}_k$, of the parameters
$(\alpha_k,x_k, y_k)$ is approximately given by\cite{Thiebaut_et_al_2018}:
\begin{equation}%
  \M{C}_k \approx \frac{1}{\eta}_k \M{J}_k\T.\M{A}_k\Inv.\M{J}_k,
\end{equation}
with  $\M{J}_k$ the Jacobian matrix of the partial derivatives of the
non-linear relationship between $\V{u}_k$ and $(\alpha_k, x_k, y_k)$.

Figure~\ref{f:wfs} shows an example of a processed frame. On top of a frame
from the wavefront sensor, pre-processed as detailed in
\refsec{s:preprocessing}, the crosses show the estimated locations $(x_k, y_k)$
of the sub-images, and the ellipses denote the estimated standard deviation
around these locations, from the coefficients of $\M{C}_k$. The ellipses have
been scaled so that the size of a sub-image correspond to the size of a pixel,
so that the standard deviation of the position is $\pm0.5$ pixel when the
diameter of the an ellipse equals the size of a sub-image. We can see
variations of the errors across the pupil, with larger errors from the
subapertures at the edge of the pupil. The ellipse are also elongated because
of the particular structure of the reference image shown in the middle of the
image: the errors are not the same along $x$ and $y$ axis, with some slight
correlation between the coordinates.

\section{Update of the reference image}
\label{s:update}

In the previous section, by assuming the reference image $\V{r}$ is known, the
optimization of the general cost fonction $\psi(\V\theta)$ given by
\refeq{e:psi_cost} allowed all the parameters but $\V{r}$ to be determined in
the set of the parameters $\V{\theta} = \{\eta_k, \alpha_k,
\V{s}_k\}_{k=1,...,n} \cup \V{r}$. In an alternate approach, we now look for
the reference image $\V{r}$ itself, assuming all the other parameters are
known.

Denoting $\M{G}_k = \alpha_k\,\M{R}_k(\V{s}_k)$ (\ie shifting the reference
image by an offset $\V{s}_k$, and rescaling by factor $\alpha_k$), we can write
the significant terms of \refeq{e:psi_cost} for our purpose as:
\begin{equation}\label{e:psi_cost_r}
  \psi^\prime(\V\theta) = \sum_{k=1}^{n}
      \,\Norm{\V{d}_k - \M{G}_k.\V{r}}_{\eta_k\M{W}_k}^2
       + \mu \Norm{\M{D}.\V{r}}_2^2.
\end{equation}

Thus the expression is reduced to regularized least-squares. The underlying
approximation here for a fast computation when applying $\M{G}_k$ is to shift
$\V{r}$ by $\V{s}_k$ rounded at the nearest pixel, thus avoiding the need of
any interpolation. 

The regularization term is mandatory here because the reference image is used
as the model to be fitted and need to be known everywhere it is needed, with a
good accuracy. So the estimation of the reference image must be extrapolated on
all the field-of-view, enlarged by the dynamic of the wavefront sensor. We
assume that the enlarged field-of-view is twice the field-of-view of the
sub-images. The chosen priors, $\M{D}$, is a difference operator that computes
finite differences of the reference image.

Minimizing \refeq{e:psi_cost_r} for $\V{r}$ yields:
\begin{equation}\label{e:r}
  \left(\sum_{k=1}^{n} \eta_k\,\M{G}_k\T\!.\M{W}_k.\M{G}_k
         +\mu\,\M{D}\T\!.\M{D}\right) \V{r}
   = \sum_{k=1}^{n} \eta_k\,\M{G}_k\T\!.\M{W}_k.\V{d}_k.
\end{equation}

This equation is solved for $\V{r}$ by using conjugate gradient method. We can
notice that the right hand side expression is the weighted sum of the
recentered sub-images.

In this process, the chicken or the egg problem is solved at the first frame by
bootstrapping the method with setting $\V{r}$ as the simple weighted average of
all the sub-images. The first set of parameters are determined with this first
reference image, up to the computation of a new reference image by solving
\refeq{e:r}. This new reference image is used for the next frame, and the first
one is deleted.

For all the following frames, a new reference image is computed as soon as the
measurements are sent to the controller that computes the commands. This new
reference is merged with the previous one by using a leaking integrator.

By merging the recentered sub-images spatially in an optimal way, the reference
image gets to be known in a field larger than the field-of-view of the
sub-images. This allows all the field-of-view of the sub-images to be used to
get the measurements of their shifts. An example of such an enlargement is
shown on \reffig{f:wfs}: the reference image is shown at the center of the
image. This reference image is much less noisy than one sub-image chosen as a
reference. Furthermore, the reference image is fixed and just evolves with the
granulation pattern, thus allowing tip/tilt to be continuously measured while
maintaining the same line of sight for a long time.

\section{Conclusion}

The presented wavefront sensing method for solar adaptive optics delivers for
each frame, both the slope measurements and their covariance matrix. By using
this information, the control loop can be robust to the fast spatial and
temporal variations of the measurement errors. The quality of the measurements
is improved by using an estimate of the reference image optimally built from
the merging of all the sub-images, thus known with more details on a larger
field-of-view and with much less noise. Furthermore, the slope measurements
obtained by fitting such a reference image on the sub-images has been shown to
be optimal\cite{Thiebaut_et_al_2018}. The bootstrapping of the algorithm has
been checked on open-loop data with various samples of the granulation.

At this time, the method is being implemented in the Real Time Computer in
order to close the loop at 1~kHz. This goal is yet to be demonstrated on sky.

\acknowledgments %
 
This project has been co-funded by the European Commission's FP7 Capacities
Programme under Grant Agreement number 312495, and the Centre National de la
Recherche Scientifique.

\bibliography{Tallon_et_al_2019_Themis_WFS_AO4ELT} %
\bibliographystyle{spiebib} %

\end{document}